\documentclass[pra, twocolumn,aps,superscriptaddress,showpacs]{revtex4-2}

\usepackage{graphicx}
\usepackage{makecell}
\usepackage{lineno}
\usepackage{amsmath,amssymb}
\usepackage[table]{xcolor}
\usepackage{ulem}
\usepackage{physics}
\usepackage[hidelinks]{hyperref}

\begin{document}

\title{Quantum logic control and entanglement in hybrid atom-molecule arrays}

\author{Chi Zhang}
\email[]{c.zhang@imperial.ac.uk}
\affiliation{Centre for Cold Matter, Blackett Laboratory, Imperial College London, Prince Consort Road, London SW7 2AZ, United Kingdom}
\author{Sara Murciano}
\affiliation{Universit\'e Paris-Saclay, CNRS, LPTMS, 91405, Orsay, France.}
\author{Nathanan Tantivasadakarn}
\affiliation{C. N. Yang Institute for Theoretical Physics,
Stony Brook University, Stony Brook, NY 11794, USA}
\author{Ran Finkelstein}
\email[]{ranf@tauex.tau.ac.il}
\affiliation{School of Physics and Astronomy, Tel Aviv University, 69978, Israel}

\begin{abstract}
Polar molecules, with their rich internal structure, offer immense potential for fundamental physics, quantum technology, and controlled chemistry. However, their utilization is currently limited because of slow and imperfect state detection and weak dipolar interaction, limiting fast and large-scale entanglement generation.
We propose and analyze a scheme for quantum logic control and measurement-based state preparation in a hybrid platform of polar molecules and neutral atoms. The method leverages fast, high-fidelity atom-molecule gates and high-fidelity atomic ancilla measurements to overcome the common challenges in molecule-only platforms, while preserving their diverse structural advantages.
The proposed atom-molecule controlled-phase gate is based on resonant dipole–dipole exchange between a molecular rotational transition and an atomic Rydberg transition, rendering it three orders of magnitude faster than any direct molecule-molecule entangling gate. 
We further study several applications of our scheme including the preparation of molecular GHZ states for quantum enhanced precision measurements, the preparation of exotic molecular qudit states with topological order, and measurement-altered criticality.
Our scheme is applicable to any polar molecule. It expands the paradigm of quantum logic control and paves the way to large-scale molecular entangled states. More generally, it highlights a concrete hybrid quantum system in which each qubit is utilized in an optimal way and where the measurement-based approach can yield a significant advantage in near-term devices.

\end{abstract}

\maketitle

\begin{table*}
\caption{\textbf{Comparison of the molecule-only platform and the hybrid atom-molecule platform for applications of molecules in fundamental physics and quantum technology.}}
\label{table_overview} 
\begin{center}
\begin{tabular}{|c|c|c|} 
\hline
 & \includegraphics[width=0.27\textwidth]{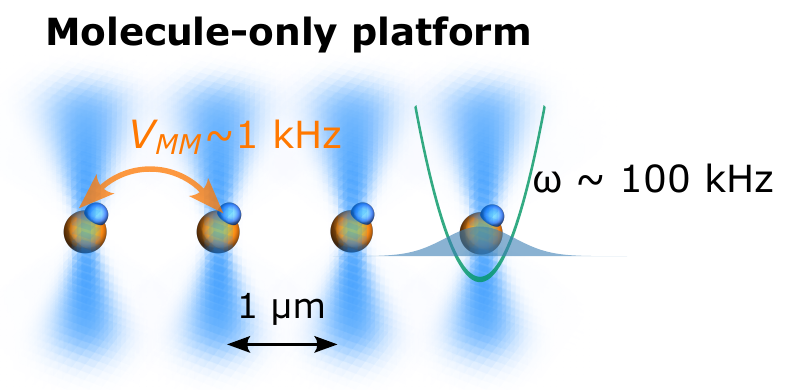} & \includegraphics[width=0.33\textwidth]{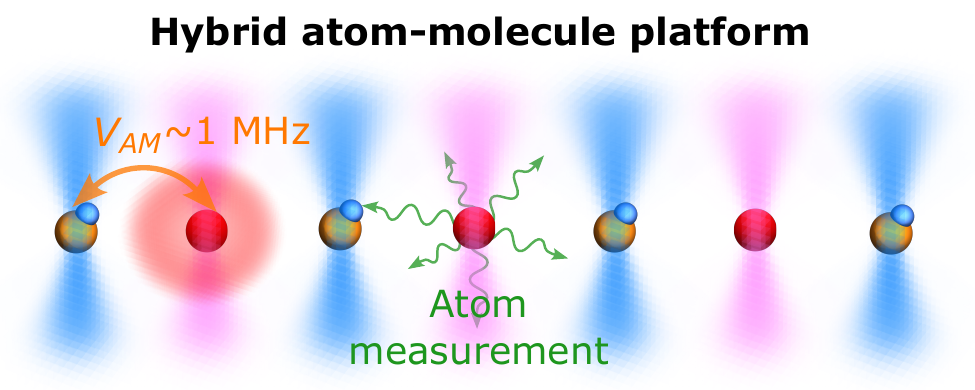} \\
\hline
\textbf{Entangling gate time} & $\gtrsim 1$~ms & $\sim 1~\mathrm{\mu s}$ \\ 
\hline
\textbf{Connectivity} & Nearest neighbor only & Long range through atoms \\ 
\hline
\textbf{Non-destructive measurement} & Depends on species & Universal through atoms \\
\hline
\textbf{Mid-circuit readout} & Low fidelity, slow (readout zone or shelving) & Fast, high-fidelity through atoms \\
\hline
\textbf{Weak measurement} & Challenging & Straightforward through atoms \\
\hline
\end{tabular}
\end{center}
\end{table*}

\section{Introduction}

Polar molecules hold great potential for a wide range of applications, including precision tests of fundamental physics \cite{Safronova2018,DeMille2024}, quantum computation and simulation \cite{Cornish2024}, and ultracold chemistry \cite{Karman2024}. Such applications uniquely benefit from the rich internal structure and versatile properties of molecules. The strong internal electric field and the high degree of control on the molecule orientation are key to recent measurements of the electron's electric dipole moment, which constrain new time-reversal symmetry violating physics to about 40~TeV \cite{Andreev2018,Roussy2024}. Molecules have various electronic, vibrational, and rotational transitions, which have different sensitivities to fundamental constants. Therefore, molecular clocks promise high-precision searches for space- and time-varying fundamental constants \cite{Truppe2013,Leung2023}. These transitions also span a broad spectrum, including telecom and microwave frequencies, offering great potential for quantum networks \cite{Lin2020}.
The complex internal structure of molecules can further be used to efficiently encode quantum information. For example, a single molecule can encode qudits or error-corrected qubits in a high-dimensional Hilbert space formed by a large number of stable states \cite{Sawant2020,Albert2020}, reducing the number of physical qubits in a quantum information processor.  

All of these applications will benefit from advanced control of molecular states and their interactions in large systems. This includes preparation and readout of single molecules, as well as generation and control of quantum correlated molecular states, such as spin squeezed states and multipartite entangled states, for quantum enhanced precision measurements and quantum simulation. 

Ultracold molecules as a quantum platform are undergoing rapid and substantial developments, including cooling into quantum degeneracy \cite{Bigagli2024,DeMarco2019}, coherence times of several seconds \cite{Gregory2021,Gregory2024}, loading reconfigurable optical tweezer arrays in motional ground states \cite{Lu2024,Bao2024}, and the realization of entangling gates \cite{Ruttley2025,Bao2023,Holland2023,Picard2025}. Compared to the neutral atom platform, molecules face several challenges; first, the state detection is relatively slow and imperfect. For molecules that can repeatedly scatter photons, the lowest state detection error of about 3\% has been achieved in 30~ms \cite{Holland2023_2}. For molecules assembled from atoms, the state detection is destructive, and its efficiency is limited by the dissociation process to about 3\% \cite{Ruttley2024}. This will limit the performance of mid-circuit operations in large systems, such as quantum error correction and scalable multipartite entanglement generation. Furthermore, the relatively weak dipole-dipole interaction limits the speed of entangling gates and thus the generation of large multipartite entangled states through unitary circuits or continuous Hamiltonian evolution. An alternative scheme to achieve long-range entanglement is to combine mid-circuit measurements with shallow unitary circuits \cite{Finkelstein2024}. However, the speed and readout efficiency of molecules limit the application of this method.

To address these challenges, here we propose a hybrid system of molecules and atoms which combines the advantageous structure of molecules for fundamental physics and quantum technology together with the pristine readout and control capabilities of atoms. The strong interaction between molecules and atoms can be used for quantum logic control of molecules using atoms. The fast and high-fidelity readout of atoms, together with the large spectral separation between atomic and molecular resonances, can enable mid-circuit ancilla-based operations, such as continuous clock~\cite{Liu2025}, dual-quadrature measurements~\cite{Shaw2024}, multi-qubit parity checks~\cite{Finkelstein2024}, and long-range entanglement generation. See Table~\ref{table_overview} for a comparison between the molecule-only platform and the hybrid platform.

The rest of the paper is organized as follows. 
In Sect.~\ref{sec_molecule}, we review the existing techniques for control, readout, and entanglement of molecules. In Sect.~\ref{sec_mol_atom}, we illustrate the basic building block of the new hybrid platform, the molecule-atom entangling gate. In Sect.~\ref{sec_GHZ}, we suggest a scheme for measurement-based generation of long-range entangled states of polar molecules which is designed to outperform schemes using only direct molecule-molecule entangling gates. As a case study, we highlight a specific application that can benefit from our proposed scheme in precision measurements and quantum metrology. Finally, in Sect.~\ref{sec_other} we discuss other applications of such hybrid systems, utilizing measurement-based schemes to prepare exotic topological order and to probe and alter critical systems.

\section{Entanglement in arrays of ultracold molecules}

\label{sec_molecule} 

Molecules have a rich internal structure, with many states suitable for encoding quantum information in their rotational and hyperfine spectrum. We choose two states labeled $\ket{1}$ and $\ket{2}$, which are connected by an electric dipole transition with a transition dipole moment $\bra{1} d \ket{2} = d_{\rm M}$, where $d$ is the dipole operator. $\ket{1}$ and $\ket{2}$ can be, for example, different rotational levels of the molecule. Two molecules separated by micrometer distances interact with each other primarily via a dipole-dipole interaction. In the $\{\ket{1}, \ket{2}\}$ basis, the interaction between two molecules is described by:

\begin{equation}
    H_{\rm MM} = \frac{V_{\rm MM}}{2}\left( \sigma^x_1 \sigma^x_2 + \sigma^y_1 \sigma^y_2 \right),
\end{equation}

where $V_{\rm MM}$ is the interaction strength between two molecules and $\sigma^{x,y}_{1,2}$ are the Pauli matrices in the $\{\ket{1}, \ket{2}\}$ basis for the first and second molecule, respectively. $H_{\rm MM}$ enables direct spin exchange interaction. 
Qubits can be encoded in the state $\ket{1}$ and another state $\ket{0}$, which does not have a transition dipole moment with $\ket{1}$. 

There are two commonly used protocols for entangling gates between molecules~\cite{Ruttley2025}. The first is the iSWAP gate, which utilizes the direct spin-exchange interaction~\cite{Cornish2024}. Two molecules are individually initialized, one in $\ket{0}$ and the other in $\ket{1}$. The state $\ket{0}$ is mapped to $\ket{2}$ and the molecules start to interact. After an interaction time of $t=\pi/(2V_{\rm MM})$, the two molecules are maximally entangled in the $\frac{1}{\sqrt{2}}(\ket{12}+i\ket{21})$ state. The interaction is turned off by mapping the state $\ket{2}$ back to $\ket{0}$. The interaction time can also be controlled by dynamically controlling the inter-molecule separation, as the interaction strength depends on distance. The exchange interaction and the iSWAP gate in particular have recently been experimentally realized~\cite{Ruttley2025,Picard2025,Bao2023,Holland2023}. 

The other type of entangling gate resonantly drives the pair-state transition between $\ket{11}\leftrightarrow\ket{\Psi^+}$, where $\ket{\Psi^+}=\frac{1}{\sqrt{2}}(\ket{12}+\ket{21})$ is an eigenstate of the dipole-dipole interaction Hamiltonian $H_{\rm MM}$. The state $\ket{11}$ undergoes a Rabi oscillation to $\ket{\Psi^+}$ and back to $-\ket{11}$ with an extra $\pi$ phase. Other pair states $\ket{00}$, $\ket{01}$, and $\ket{10}$ are off-resonant with the driving field due to the energy shift of $\ket{\Psi^+}$ by dipole-dipole interaction. This realizes a controlled phase gate and has been experimentally demonstrated in RbCs molecules \cite{Ruttley2025}.  
\begin{figure*}
	\includegraphics[width=0.8\textwidth]{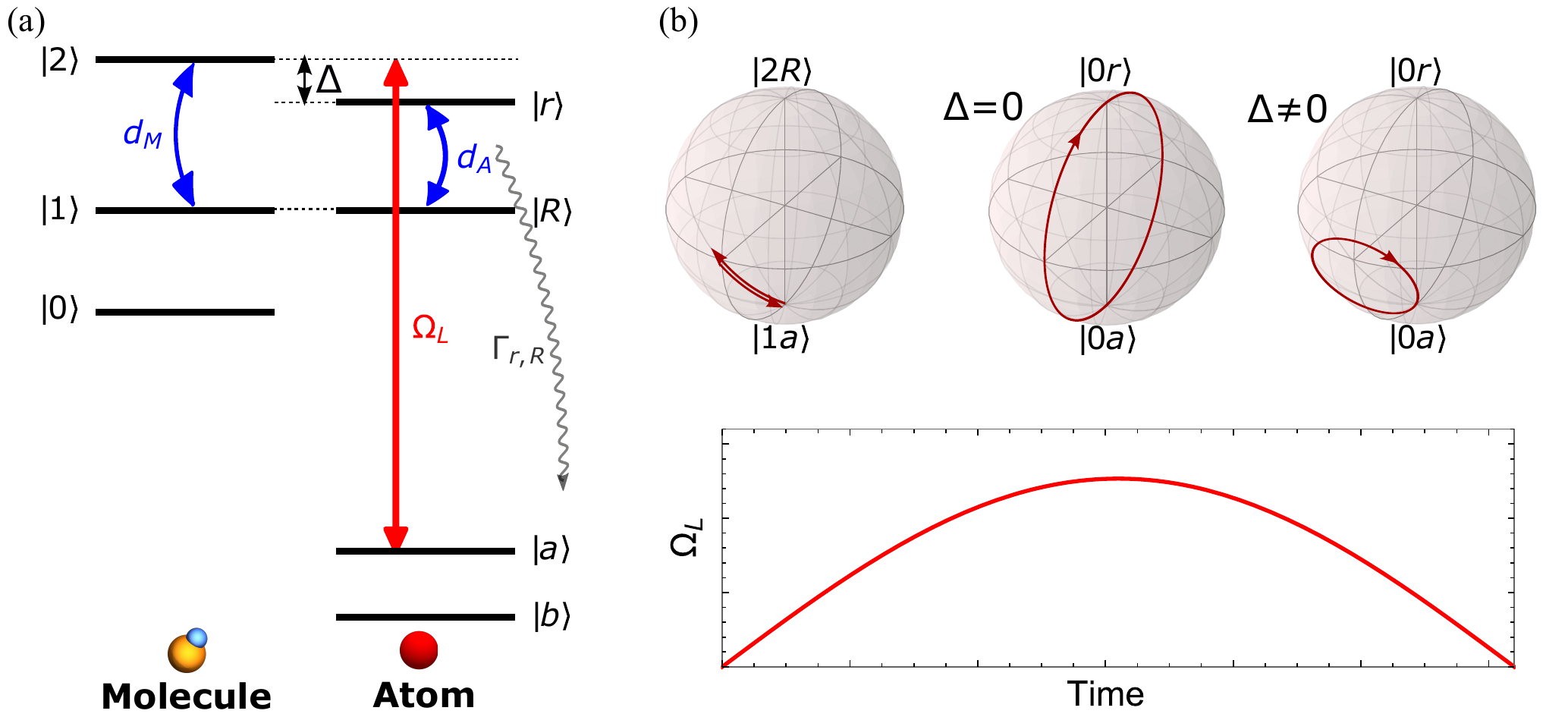}
	\caption{\textbf{A controlled-phase gate between a molecule and an atom.} \textbf{(a)} $\ket{0}$, $\ket{1}$ and $\ket{2}$ are molecular states, $\ket{1}$ and $\ket{2}$ are connected with an electric dipole transition. $\ket{a}$ and $\ket{b}$ are low-lying atomic states. $\ket{r}$ and $\ket{R}$ are atomic Rydberg states which are connected with an electric dipole transition. The transition frequency of $\ket{r}\leftrightarrow\ket{R}$ can be tuned by an electric field to near resonance with $\ket{1}\leftrightarrow\ket{2}$. A laser field couples $\ket{a}\leftrightarrow\ket{r}$. \textbf{(b)} The gate sequence consists of a sinusoidal-shaped laser pulse applied to the atom. When $\Delta=0$ and the pulse area is $2\pi$, it is a controlled Z gate. The maximum coupling strength $\Omega_\mathrm{L,max} \lesssim V_\mathrm{MA}$. For the pair state initially in $\ket{0a}$, the system acquires a $\pi$ phase after a Rabi oscillation to $\ket{0r}$ (middle Bloch sphere). For $\ket{1a}$, the system follows the eigenstate adiabatically and returns to $\ket{1a}$ without an extra phase (left Bloch sphere). $\ket{0b}$ and $\ket{1b}$ are not coupled by the laser and stay in the same states. When $\Delta\neq 0$ it is a controlled arbitrary phase gate. For the pair state initially in $\ket{0a}$, by choosing an appropriate pulse shape, the system undergoes an off-resonant Rabi oscillation and returns to $\ket{0a}$ with a closed loop on the Bloch sphere (right Bloch sphere). For other initial states, it is similar to the $\Delta=0$ case.}
	\label{Fig_atom_mol}
\end{figure*}
The gate time for entangling two molecules, in both protocols, is fundamentally limited by the interaction strength, which is between 100~Hz and 1~kHz for molecules with dipole moments of a few Debye separated by 2~$\mathrm{\mu m}$. Additionally, the dipole-dipole interaction is distance-dependent, and thus couples the qubit states to molecular motion. This unwanted coupling causes decoherence, as well as heating which in turn leads to further decoherence and further heating. Molecules can be trapped with much shorter separations ($\sim$100~nm) using state-dependent optical tweezers \cite{Caldwell2020c} for enhancing the interaction strength. However, the decoherence due to motion is also enhanced for smaller molecule separations, since the spread of the motional wavefunction causes larger variation of the interaction strength, limiting the performance of fast entangling gates.

To scale up the number of entangled molecules for quantum metrology and quantum simulation applications, a large number of two-molecule gates will need to be applied. The slow gate time will limit the circuit depths due to the finite coherence time. Additionally, heating due to unwanted motional coupling may limit the number of successive gates.
Alternatively, long-range multipartite entanglement can be prepared using fixed-depth, shallow circuits with techniques borrowed from measurement-based quantum computation, which require high-fidelity mid-circuit readout~\cite{Raussendorf05,Piroli2021_Circuits,Verresen2021,Tantivasadakarn24,Lee2022,Zhu23,Nat2023_hierarchy}. In trapped molecules such as CaF and RbCs, this will be primarily limited by the readout error of 3\%~\cite{Holland2023_2,Ruttley2024}. In contrast, in the next section we show that the measurement-based approach can be highly efficient in a hybrid platform of atoms and molecules.

\section{Atom-molecule gate for quantum logic operations}
\label{sec_mol_atom}

To address the challenges in entanglement and readout of molecules, we suggest a hybrid system of atoms and molecules, where inter-species gates can be much faster and readout error can be made much lower.
Quantum logic spectroscopy is a powerful technique initially developed in ion traps \cite{Tan2015,Lin2020}. It has enabled readout and control of the quantum state of target ions using an ancillary ion without direct readout and photon scattering of the target ion. The method relies on the phonon-mediated interaction between different ion species and works generically for any ion species, including molecular ions \cite{Lin2020}. This method paved the way for a broad range of applications in ion traps from atomic clocks \cite{Marshall2025}, to precision tests of fundamental physics \cite{Zhou2024}, and quantum information processing \cite{Tan2015}. 

Here we propose a quantum logic spectroscopy scheme for neutral molecules using Rydberg atoms. The key ingredients of our method are the entangling gate between a molecule and an atom, and the readout based on the atomic ancilla qubit. We further extend the idea beyond spectroscopy to measurement-induced generation of long-range entangled states.
Fig.~\ref{Fig_atom_mol} shows the basic idea. We choose two atomic states $\ket{a}$ and $\ket{b}$ to encode the atomic qubit. $\ket{a}$ is coupled to a Rydberg state $\ket{r}$ by a laser field $\Omega_{\rm L}$. Another Rydberg state $\ket{R}$ is coupled to $\ket{r}$ via a transition dipole moment $d_{\rm A}$. The transition between $\ket{r}\leftrightarrow\ket{R}$ can be tuned to resonance with the molecular transition $\ket{1}\leftrightarrow\ket{2}$ by a small electric field. The pair states $\ket{1r}$ and $\ket{2R}$ are then coupled by the dipole-dipole interaction

\begin{equation}
    H_{\rm MA} = \frac{V_{\rm MA}}{2}\left( \ket{1r}\bra{2R} + \ket{2R}\bra{1r} \right) + \Delta \ket{2R} \bra{2R},
\end{equation}

where $\Delta$ is the detuning between the pair states $\ket{1r}$ and $\ket{2R}$. Typically $V_{\rm MA}$ is at least three orders of magnitude larger than $V_{\rm MM}$ at similar separations, due to the large dipole moment of Rydberg states ($d_A \gtrsim 1000 d_M$).
It can thus yield an effect similar to the Rydberg blockade, where an atomic excitation can be suppressed conditioned on the state of the molecule \cite{Zhang2022b}. A pulse sequence which turns this into a molecule-atom controlled-Z gate is shown in Fig.~\ref{Fig_atom_mol}. We tune the transitions $\ket{r}\leftrightarrow\ket{R}$ and $\ket{1}\leftrightarrow\ket{2}$ to resonance ($\Delta=0$). A sinusoidal-shaped laser pulse with a maximal Rabi frequency $\Omega_\mathrm{L,max} \lesssim V_\mathrm{MA}$ and a pulse area of $2\pi$ is applied to the atom. The pair state $\ket{0a}$ acquires a geometrical $\pi$ phase after a Rabi oscillation through $\ket{0r}$. For $\ket{1a}$, the system follows the eigenstate adiabatically and returns to $\ket{1a}$ without an extra phase. $\ket{0b}$ and $\ket{1b}$ are not coupled by the laser and remain unchanged. This realizes a controlled-Z gate between the atom and the molecule.

The same sequence can also be used to implement a general, arbitrary controlled-phase gate. Using electrical field control, the transition frequency of $\ket{r}\leftrightarrow\ket{R}$ is shifted from $\ket{1}\leftrightarrow\ket{2}$ by $\Delta\lesssim V_\mathrm{MA}$. The $\ket{a}\leftrightarrow\ket{r}$ coupling laser is also detuned by $\Delta$. The gate sequence is still a smooth-varying laser pulse. For an initial two-body state $\ket{0a}$, the Rabi oscillation is off-resonant. An arbitrary phase can be realized by choosing an appropriate pulse area. This implements a controlled arbitrary phase gate.

\begin{table*}
\caption{\textbf{Gate error and scaling with molecule properties.} The interaction strength $V_\mathrm{MA}\propto d_Ad_M \propto n^{2} d_M$, where $n$ is the principal quantum number. The choice of Rydberg states depends on molecule species; the Rydberg level spacing $\propto n^{-3}$ needs to roughly match the rotational frequency $f$ of the molecule so that they can be tuned to resonance with a small electric field. Laser coupling strength is typically a few MHz, and is chosen to optimize decay error and non-adiabatic error (see Ref.~\cite{Zhang2022b}). As a result, all these errors depend on the rotational frequency and the dipole moment of the molecule. The source of detuning is assumed from Stark shift of Rydberg atoms in an electric field fluctuation of $\sim 0.1~\mathrm{mV/cm}$.}
\label{table1} 
\begin{center}
\begin{tabular}{ |c|c| c| } 
\hline
Error source & Scaling & Error for CaF and Rb\\
\hline
Rydberg state decay  & $f^{3/2} d_M^{-1}$ & $7\times10^{-4}$ \\ 
\hline
Adiabaticity breakdown  & - & $2.5\times10^{-4}$ \\ 
\hline
Leakage out of qubit subspace & $f^{-10/3}d_M^2$ & $5\times 10^{-8}$ \\ 
\hline
Electrical field fluctuations  & $f^{10/3} d_M^{-2}$ & $8\times 10^{-5}$ \\
\hline
Total error & - & $1\times10^{-3}$ \\
\hline
\end{tabular}
\end{center}
\end{table*}

\subsection*{Atom-molecule gate error budget}

The atom-molecule gate error sources are listed and calculated for CaF and Rb in Table~\ref{table1}. The mechanism of this gate is similar to that of the two-molecule gate mediated by Rydberg atoms in Ref.~\cite{Zhang2022b}. Therefore, the gate error is calculated using the same method. There are multiple choices of atomic and molecular states for the gate. For concreteness, here we choose \mbox{$\ket{0}=\ket{N=1,F=0,M=0}$} ($N,F,M$ are rotational, hyperfine, and magnetic quantum numbers, respectively), \mbox{$\ket{1}=\ket{N=0,F=0,M=0}$}, and {$\ket{2}=\ket{N=1,F=1^-,M=1}$} (the superscript in $F$ indicates the parity of the state) for CaF, and \mbox{$\ket{r}=\ket{59s_{1/2},F=1,M=-1}$} and $\ket{R}=\ket{58p_{3/2},F=2,M=-2}$ for Rb. It is important to note that the gate error is highly independent of the specific choice of states and depends primarily on the rotational constant and dipole moment of the molecule. The main error sources include Rydberg state decay, non-adiabatic transitions for state $\ket{1a}$, unwanted interactions from other atomic and molecular states that are outside of the computational subspace, and static detuning between the molecule and Rydberg transitions caused by imperfect control of the electric field.

\section{Measurement-based preparation of large-scale molecular GHZ states}
\label{sec_GHZ}
We now turn to show how such molecule-atom gates can be used to prepare large entangled states of molecules in a scheme which has the potential to be more efficient than a quantum unitary circuit with direct molecule-molecule gates only. 
We suggest utilizing the combination of high-fidelity atom-molecule gates introduced in Sect.~\ref{sec_mol_atom} with the high-fidelity state detection of atoms (as compared with molecules) to scale up entangled states of molecules. We focus on states which are advantageous in quantum sensing and in probing fundamental physics, for instance the Greenberger-Horne-Zeilinger (GHZ) state that can enhance the measurement precision of fundamental symmetry violations in molecules beyond the standard quantum limit while exhibiting reduced sensitivity to external field noise~\cite{Zhang2023}. 

\begin{figure*}
	\includegraphics[width=0.6\textwidth]{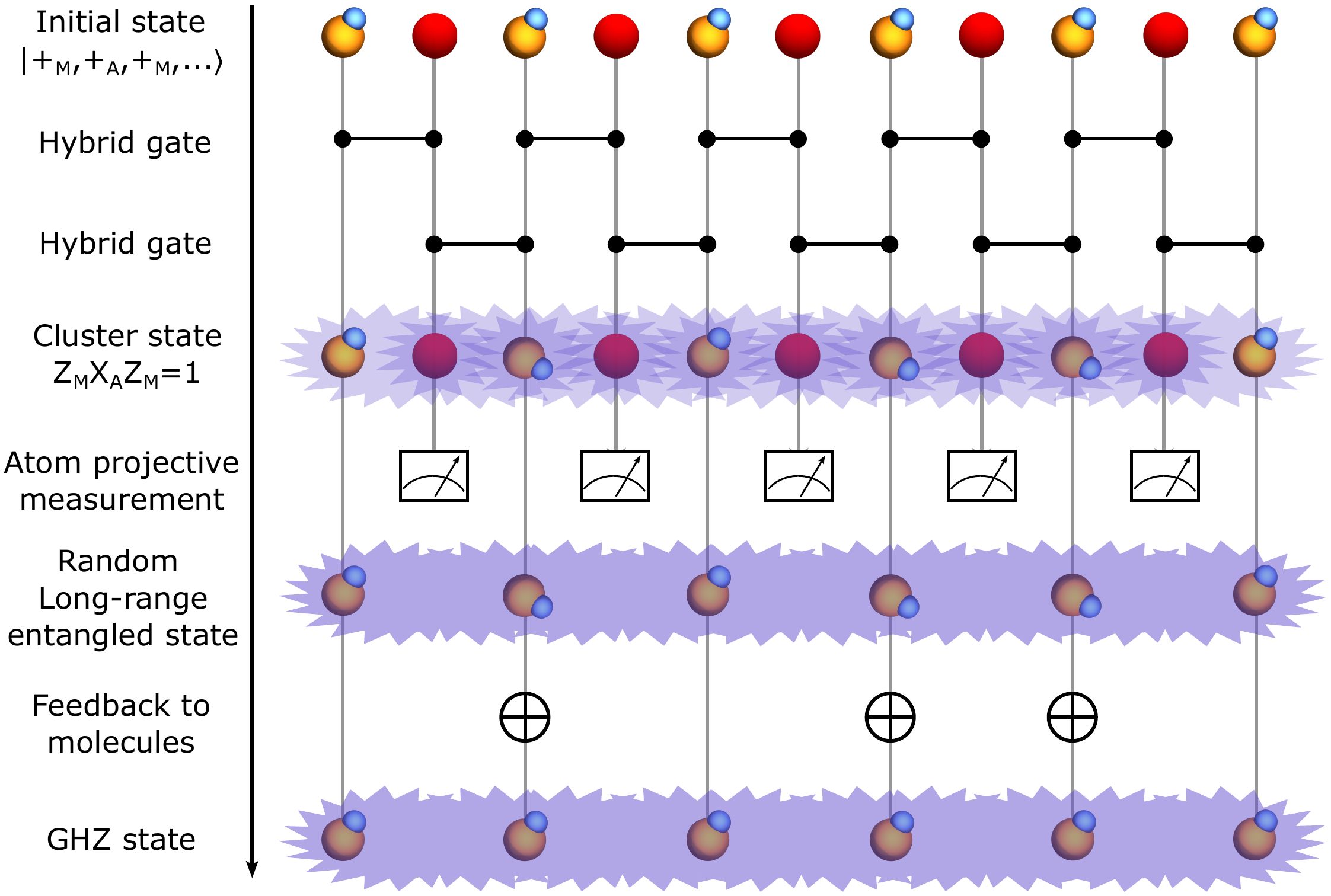}
	\caption{\textbf{Atomic ancilla measurement-based generation of a long-range entangled state of molecules.} All atoms and molecules are initialized in $\ket{+}$ (for atoms $\ket{\pm} = \frac{1}{\sqrt{2}} (\ket{0} \pm \ket{1})$, for molecules $\ket{\pm} = \frac{1}{\sqrt{2}} (\ket{a} \pm \ket{b})$). Atoms are first moved to entangle with the molecules on their left side and then with the molecules on their right side. A cluster state is generated. Subsequently, all atoms are measured in the $\ket{\pm}$ basis. The subsystem of molecules is projected to a long-range entangled state, which is correlated with the states of the atoms. Based on the measurement results of the atoms, the molecular state can be rotated to the target long-range entangled state (e.g. the GHZ state) by single molecule rotations.}
	\label{Fig_GHZ}
\end{figure*}

Our proposal is based on a well-known scheme in the field of measurement-based quantum computation~\cite{Briegel01}, which has recently gained significant interest in the context of long-range entangled state preparation in matter-based quantum processors with mid-circuit measurement capabilities~\cite{Raussendorf05,Piroli2021_Circuits,Verresen2021,Tantivasadakarn24,Lee2022,Zhu23,Nat2023_hierarchy}. 
The scheme (Fig.~\ref{Fig_GHZ}) is composed of generating a one-dimensional cluster state using a shallow-depth circuit and performing measurements in the $X$-basis on the atomic sub-lattice. The remaining unmeasured molecular qubits are then projected to a well-defined cat state which is correlated with the measurement outcome of the atoms and can be turned into the GHZ state by performing single-qubit rotations based on the atomic measurement results as a feedforward layer. Qubits can be selected by focused two-photon laser beams or by a global microwave field together with local light shifts induced by varying the tweezer power.

This scheme offers a clear shortcut to preparing long-range entangled states, specifically the GHZ state which is an optimal input state for local phase estimation with applications in quantum metrology. 
However, the state preparation fidelity for this scheme depends on the fidelity of mid-circuit measurements, on the cluster state preparation fidelity, and on the coherence time of the molecular qubits.
These should be compared with the fidelity of a larger depth circuit where the number of gates scales as $\mathrm{N}$ or $\mathrm{log(N)}$ for gate-only schemes. 
We identify our suggested system as a platform where such a measurement-based approach is especially suitable and may yield an advantage in the near term, as the inherent advantages of the hybrid atom-molecule system naturally match the measurement-based scheme.

As discussed in Sect.~\ref{sec_mol_atom}, while direct molecule-molecule gates have made great progress, they are still limited in speed (due to relatively low interaction energy), which typically limits their fidelity. Additionally, reading out the molecule state with high fidelity is a long-standing challenge. 
A hybrid atom-molecule system, on the other hand, would leverage the advantages of atoms such as high-fidelity fast readout and high-fidelity, fast hybrid atom-molecule gates, to improve molecule GHZ state preparation.

\section{Other applications of hybrid atom-molecule gates}
\label{sec_other}

\subsection*{Utilizing higher dimensional encodings to prepare exotic topological order}
Molecules host a wide range of stable levels, rendering them as natural candidates to encode quantum information using d-levels or qudits. In the context of measurement-induced phases of matter, it is particularly relevant to consider 3-level systems or qutrits. The realization of a qutrit entangling gate between an atom and a molecule can follow as an extension of the atom-molecule gate we have presented in Sect.~\ref{sec_mol_atom}. This extension is based on a recently proposed scheme for a qutrit-entangling gate in neutral atoms, which is composed from a pulse sequence of alternating controlled-phase gates applying up to two Rydberg levels and single qutrit rotations~\cite{Burshtein2025}.

Thus, achieving qutrit gates on our proposed platform requires only an additional stable level in both the molecule and the atom, and the ability to perform single qutrit rotations on each species, which can also be easily achieved by adding one more coupling field to the newly introduced level.
Using qutrit encoding with atomic ancilla mid-circuit measurement can be used for efficient preparation of the ground state of the qutrit GHZ state as well as the $\mathbb{Z}_3$ Toric code through measurement and feedforward on the qutrit cluster state.

The $\mathbb{Z}_3$ Toric code can be further utilized as a resource state to prepare the $S_3$ quantum double model.
$S_3$ is the smallest non-abelian group and enables universal topologically protected quantum computation. The measurement-based generation of such a topological order requires the ability to perform a gate between the $\mathbb{Z}_3$ cluster state and a second set of qubits prepared in the $\mathbb{Z}_2$ cluster state, followed by a round of measurement and feedforward~\cite{Verresen2021,Nat2023_hierarchy,Lo2026universal}.
We note that such an application combined the inherent advantages of qudit encodings in molecules, while overcoming the challenges in molecular systems such as low fidelity measurement and direct entangling gates. Thus, we foresee that this type of application will be a promising setup to observe true advantage in measurement-based state preparation and in generating such non-abelian topological order. 



%

\subsection*{Measurement-altered criticality}

Hybrid platforms that combine molecules and atoms offer a promising route for implementing weak measurements on the ground state of many-body quantum systems~\cite{garratt2023}. Such systems would be encoded in molecular degrees of freedom, which allow for the engineering of a wide variety of spin models. For instance, the XXZ spin chain model,
\begin{equation}\label{eq:xxz}
    H=-\sum_j [X_j X_{j+1}+Y_j Y_{j+1}+\Delta Z_j Z_{j+1}],
\end{equation}
can be realized using dipolar interactions, as shown in Ref. \cite{Micheli2006,Christakis2023}. Beyond qubit-based models, molecular systems can also access larger local Hilbert space dimensions. For instance, the three-state Potts model can be engineered using qutrits encoded in three internal states of a molecule, with the Hamiltonian
\begin{equation}\label{eq:Potts}
    H=-\sum_j\left(J U_j U^{\dagger}_{j+1}+J U^{\dagger}_j U_{j+1}+h(V_j+V_j^{\dagger})\right),
\end{equation}
where the on-site operators are
\begin{equation}
    V_j=\begin{pmatrix}
    1 & 0 & 0 \\
    0 & e^{i 2\pi/3} & 0 \\
    0 & 0 & e^{i 4\pi/3}
    \end{pmatrix}, ~~U_j=\begin{pmatrix}
    0 & 0 & 1 \\
    1 & 0 & 0 \\
    0 & 1 & 0
    \end{pmatrix}.
\end{equation}
This Hamiltonian can be realized in a molecule qutrit system by dressing three rotational states with microwave fields such that there are equal-strength dipole-allowed transitions between all three.
By tuning the ratio $J/h$, the ground state of this Hamiltonian interpolates between an ordered and a disordered $Z_3$ symmetry-broken phase. We are interested in the self-dual point $J=h$, where the system undergoes a continuous phase transition in the three‑state Potts universality class~\cite{Belavin1984}. 
In such a setup, atoms serve as ancillary systems that are more easily prepared and measured. This enables weak measurement protocols via the following scheme:
\begin{enumerate}
    \item \textbf{Preparation}: Initialize molecules in the ground state of a critical spin model, such as the ground state of Eq.~\eqref{eq:xxz} or \eqref{eq:Potts}).
    Ground state preparation can be realized in an analog simulator setting by an adiabatic sweep of the Hamiltonian parameters. Alternatively, the ground states of Eqs.~\eqref{eq:xxz} and \eqref{eq:Potts}) can be prepared with a linear depth circuit, using direct molecule-molecule gates~\cite{Ruttley2025,Picard2025,Holland2023} or gates mediated by atoms~\cite{Zhang2022b}. The digital circuit approach can include the use of variational optimization algorithms and explicit optimized circuits for each size of the system \cite{Ho19,Rogerson24}. 
    Atoms are initialized in a state that can be a simple product state or an entangled state. 
    \item \textbf{Entangling gate}: Apply a controlled-phase gate on atom-molecule pairs following the protocol in Sect.~\ref{sec_mol_atom}. 
    The regime of weak measurements is achieved by applying small controlled-phase gates, which we expect to be realized with very high fidelity as their duration can be very short.
    \item \textbf{Measurement}: Once atoms and molecules are correlated, measuring the atomic degree of freedom then induces an effective measurement on the molecular system. One can bias the measurement statistics to increase the success probability of post-selecting a desired outcome by preparing the ancilla system accordingly. 
\end{enumerate}

To give more intuition about the last point, for example, in Ref.~\cite{Murciano2023}, both the critical system and the ancilla system are taken to be transverse‑field Ising models. By choosing the ancilla to be in its paramagnetic phase, post‑selection of the uniform (most likely) measurement outcome can be carried out with a surprisingly modest number of repetitions even at large system sizes. This efficiency arises because the measurement outcomes of the ancilla follow a strongly biased probability distribution. 

Recent proposals have shown the possibility of observing measurement-altered criticality in single Rydberg chains without ancillas (e.g., via partial measurement protocols), enhancing the post-selection probability~\cite{Naus2025}. The main advantage of molecule-atom architectures is that they can expand the range of accessible models and measurement channels. In particular, even though finding the optimal measurement scheme for critical theories beyond the Ising case will require further study, the use of entangled ancillas opens the door to engineering post-selected states with long-range correlations and tunable entanglement. Preparing the Potts critical point would open a route to experimentally probing connections between measurement-induced effects in many-body states and nontrivial renormalization-group flows~\cite{yueliu2025}.

\section{Summary and Outlook}

We have proposed a hybrid platform of molecules and atoms to significantly advance quantum logic control and entanglement of polar molecules. In this system, molecules can be used as quantum sensors for fundamental physics, qubits and qudits for quantum simulation, and elementary ingredients for controlled chemistry. Rydberg atoms interact with molecules through the dipole-dipole interaction. This interaction enables fast, high-precision control and mid-circuit readout of the internal states of molecules. It also advances the manipulation of many-body molecular systems; for example, long-range entanglement can be generated by mid-circuit operations, and many-body correlations can be probed by ancilla-based weak measurements. This platform combines the inherent complexity and versatility of molecules with the high fidelity control, measurement, and entanglement of atoms. Notably, co-trapping of various species of molecules and atoms has been experimentally realized recently~\cite{Guttridge2023,Zhu2025,Jurgilas2021,Jurgilas2021b}. Our work can guide the design of such near-term experiments as it is broadly applicable to all kinds of polar molecules, and will advance various applications in molecular quantum technology in the near future.

\textit{Note --} During completion of this work we became aware of a relevant work on non-destructive measurement of molecules using Rydberg atoms~\cite{Young2026}.
\begin{acknowledgements}
We thank Manuel Endres and Nick Hutzler for insightful suggestions and discussions at the early stages of this work. We thank Lewis Picard for valuable feedback on this manuscript.
We acknowledge support from the Walter Burke Institute for Theoretical Physics at Caltech.
This research was supported by Israel Science Foundation research grant (ISF’s No. 4098/25) and the Maimonides Fund’s Future Scientists Center.
RF acknowledges support from the Troesh postdoctoral fellowship while working at Caltech. CZ was supported by the Caltech David and Ellen Lee Postdoctoral Fellowship.
SM thanks Jason Alicea, Yue Liu, Nandagopal Manoj, Stephen Naus, Pablo Sala for discussions and collaborations on related topics.

\end{acknowledgements}

%
\end{document}